\documentclass[preprintnumbers,amsmath,amssymbm,preprint]{revtex4}
\usepackage{epsfig}
\usepackage{graphicx}

\begin{document}
\title{Marginally bound circular orbits in the composed black-hole-ring system}
\author{Shahar Hod}
\address{The Ruppin Academic Center, Emeq Hefer 40250, Israel}
\address{ }
\address{The Hadassah Institute, Jerusalem 91010, Israel}
\date{\today}

\begin{abstract}
The physical and mathematical properties of the non-linearly coupled
black-hole-orbiting-ring system are studied analytically to second
order in the dimensionless angular velocity
$M_{\text{ir}}\omega_{\text{H}}$ of the black-hole horizon (here
$M_{\text{ir}}$ is the irreducible mass of the slowly rotating
central black hole). In particular, we determine analytically, to
first order in the dimensionless ring-to-black-hole mass ratio
$m/M_{\text{ir}}$, the shift
$\Delta\Omega_{\text{mb}}/\Omega_{\text{mb}}$ in the orbital
frequency of the {\it marginally bound} circular geodesic that
characterizes the composed curved spacetime. Interestingly, our
analytical results for the frequency shift
$\Delta\Omega_{\text{mb}}$ in the composed black-hole-orbiting-ring
toy model agree qualitatively with the recently published numerical
results for the corresponding frequency shift in the physically
related (and mathematically much more complex)
black-hole-orbiting-particle system. In particular, the present
analysis provides evidence that, at order $O(m/M_{\text{ir}})$, the recently
observed positive shift in the angular frequency of the marginally
bound circular orbit is directly related to the physically intriguing
phenomenon of dragging of inertial frames by orbiting masses in
general relativity.
\end{abstract}
\maketitle

\section{Introduction}

Geodesic orbits of test particles in curved spacetimes are of
central importance in black-hole physics
\cite{Car,Bar,Chan,Shap,WT,Willmb,Gro,Hodmb}. They provide valuable
information on the physical parameters (mass, charge, angular
momentum) of the central black hole. In particular, the {\it
marginally bound} circular orbit of a curved black-hole spacetime is
of special importance in astrophysics and general relativity 
\cite{Car,Bar,Chan,Shap,WT,Willmb,Gro,Hodmb}. This physically
interesting geodesic represents the innermost circular orbit of a
massive particle which is energetically bound to the central black
hole.

For a test particle of proper mass $m$, the marginally bound
circular geodesic is characterized by the simple energy relation
\cite{Car,Bar,Chan,Shap}
\begin{equation}\label{Eq1}
E(r_{\text{mb}})=m\  ,
\end{equation}
where $E$ is the energy of the particle as measured by asymptotic
observers. Interestingly, the marginally bound circular geodesic
(\ref{Eq1}) marks the boundary between bound orbits, which are
characterized by the sub-critical energy relation $E<m$, and unbound
circular orbits with $E>m$ which, given a small outward
perturbation, can escape to infinity. In particular, as nicely
demonstrated in \cite{Willmb,WT}, the critical (marginally bound)
circular geodesic (\ref{Eq1}) plays a key role in the dynamics of
star clusters around super-massive black holes in galactic nuclei.
[The critical orbit (\ref{Eq1}) is sometimes referred to in the
physics literature as the innermost bound spherical orbit (IBSO)
\cite{Willmb,Gro}].

An important gauge-invariant physical quantity that characterizes
the motion of particles along the marginally bound circular geodesic 
is the orbital angular frequency $\Omega_{\text{mb}}$ of the
particles as measured by asymptotic observers. For a test-particle
moving in the spinless spherically symmetric Schwarzschild
black-hole spacetime, this physically important orbital frequency is
given by the simple dimensionless relation \cite{Bar,Chan,Shap}
\begin{equation}\label{Eq2}
M_{\text{ir}}\Omega_{\text{mb}}={1\over 8}\  .
\end{equation}
Here $M_{\text{ir}}$ is the irreducible mass \cite{Noteirr} of the
central black hole.

In recent years, there is a growing physical interest in calculating
the $O(m/M_{\text{ir}})$ corrections to the orbital frequency
(\ref{Eq1}) of the marginally bound circular orbit in non-linearly
coupled black-hole-particle systems (see the physically interesting
work \cite{Barackmb} and references therein). To this end, one
should take into account the gravitational self-force corrections to
the geodesic orbits of the particles 
\cite{Ori,Poi,Lou1,Det1,Bar1,Det2,Sag,Kei,Sha,Dam,Bar2,Fav2,Kol}.

The gravitational self-force  has two distinct physical
contributions to the dynamics of a composed black-hole-particle
system:
\newline
(1) It is responsible for non-conservative physical
effects, like the emission of energy and angular momentum in the
form of gravitational waves \cite{Ori}.
\newline
(2) The composed black-hole-particle system is also characterized by
conservative gravitational self-force effects that preserve the
total energy and angular momentum of the system but shift the
orbital frequency of the marginally bound orbit.

Computing the gravitational self-force (GSF) correction
$\Delta\Omega_{\text{mb}}$ to the zeroth-order frequency (\ref{Eq1})
of the marginally bound circular orbit is a highly non-trivial task.
Intriguingly, Barack at. al. \cite{Barackmb} have recently used
sophisticated numerical techniques in the composed
Schwarzschild-black-hole-orbiting-particle system in order to
compute the characteristic shift $\Delta\Omega_{\text{mb}}$ in the
orbital frequency of the marginally bound orbit which is caused by
the conservative part of the GSF.

In particular, Barack at. al. \cite{Barackmb} have found the ({\it
positive}) dimensionless value
\begin{equation}\label{Eq3}
{{\Delta\Omega_{\text{mb}}}\over{\Omega_{\text{mb}}}}=c\cdot \eta+O(\eta^2)\ \ \ \text{with}\ \ \
c\simeq 0.5536\
\end{equation}
for the shift in the orbital frequency of the marginally bound
circular orbit, where
\begin{equation}\label{Eq4}
\eta\equiv {{m}\over{M_{\text{ir}}}}\
\end{equation}
is the dimensionless ratio between the mass of the orbiting particle
and the mass of the central Schwarzschild black hole. The physical
importance of the result (\ref{Eq3}) of \cite{Barackmb} stems from
the fact that it provides gauge-invariant information about the
strong-gravity effects in the highly curved region ($r\simeq
4M_{\text{ir}}$) of the black-hole spacetime.

The main goal of the present compact paper is to use {\it
analytical} techniques in order to gain some physical insights on
the intriguing $O(m/M_{\text{ir}})$ {\it increase} [see Eq.
(\ref{Eq3})] in the orbital frequency of the marginally bound
circular orbit as recently observed numerically in the physically
important work \cite{Barackmb}. In particular, we shall analyze a
simple black-hole-orbiting-ring toy model which, as we shall
explicitly show below, captures some of the essential physical
features of the (astrophysically more interesting and mathematically
much more complex) black-hole-orbiting-particle system in general
relativity. As nicely proved by Will \cite{Will}, the composed
black-hole-orbiting-ring toy model is amenable to a perturbative
analytical treatment to second order in the dimensionless angular
velocity $M_{\text{ir}}\omega_{\text{H}}$ of the central slowly
rotating black hole.

\section{The orbital frequency of the marginally bound circular orbit in
the composed black-hole-orbiting-ring spacetime}

In the present paper we would like to gain some {\it analytical}
insights into the conservative part of the $O(m/M)$-shift in the
orbital frequency $\Omega_{\text{mb}}$ of the marginally bound orbit
that has recently been computed {\it numerically} in the highly
interesting work \cite{Barackmb}. To this end, we shall use the
analytically solvable model of an axisymmetric ring of matter which
orbits a central slowly spinning black hole \cite{Will}. In
particular, we shall use this simplified axisymmetric toy model 
(which, due to its symmetry, has no dissipative effects) in order to
model the conservative part of the dynamics of the mathematically
more complex black-hole-orbiting-particle system \cite{Notengw}.

We expect the composed axisymmetric black-hole-orbiting-ring system
to capture, at least qualitatively, the essential physical features
that characterize the conservative dynamics of the composed
black-hole-orbiting-particle system. In particular, both the
orbiting particle in the black-hole-particle system and the orbiting
ring in the black-hole-ring system drag the generators of the
central black-hole horizon \cite{Will}.

The physically intriguing general relativistic effect of dragging of
inertial frames by an orbiting object is reflected, both in the
black-hole-particle system and in the black-hole-ring system, by a
non-linear spin-orbit interaction term of order
$\omega_{\text{H}}\cdot j$ in the total gravitational energy of the
composed systems (here $\omega_{\text{H}}$ is the angular velocity
of the black-hole horizon and $j$ is the angular momentum per unit
mass of the orbiting ring).

Interestingly, and most importantly for our analysis, the main
mathematical advantage of the black-hole-orbiting-ring system over
the physically more interesting (but mathematically more complex)
black-hole-orbiting-particle system stems from the fact that the
spin-orbit interaction term in the black-hole-ring system is known
in a closed {\it analytical} form to second order in the
dimensionless angular velocity $M_{\text{ir}}\omega_{\text{H}}$ of
the central black hole \cite{Will} [see Eq. (\ref{Eq10}) below].

In a very interesting work, Will \cite{Will} has analyzed the total
gravitational energy and the total angular momentum of a stationary
physical system which is composed of an axisymmetric ring of
particles of proper mass $m$ which orbits a central slowly rotating
black hole of an irreducible mass $M_{\text{ir}}$. In particular, it
has been proved in \cite{Will} that the composed axisymmetric
black-hole-orbiting-ring system is characterized by the total
angular momentum
\begin{equation}\label{Eq5}
J_{\text{total}}(x)=mj+4M^3_{\text{ir}}\omega_{\text{H}}-8mjx^3\ ,
\end{equation}
where
\begin{equation}\label{Eq6}
x\equiv {{M_{\text{ir}}}\over{R}}\
\end{equation}
is the dimensionless ratio between the irreducible mass of the
black hole and the proper circumferential radius of the ring,
\begin{equation}\label{Eq7}
j(x)={{M_{\text{ir}}}\over{[x(1-3x)]^{1/2}}}\cdot[1+O(M_{\text{ir}}\omega_{\text{H}})]\
\end{equation}
is the angular momentum per unit mass of the orbiting ring, and
$\omega_{\text{H}}$ is the angular velocity of the black-hole
horizon.

Since the first term on the r.h.s of (\ref{Eq5}) represents the
angular momentum $J_{\text{ring}}$ of the orbiting ring of mass $m$,
one concludes \cite{Will} that the last two terms in (\ref{Eq5})
represent the angular momentum
\begin{equation}\label{Eq8}
J_{\text{H}}=4M^3_{\text{ir}}\omega_{\text{H}}-8mjx^3\
\end{equation}
which is contributed by the slowly spinning central black hole as
measured by asymptotic observers. In particular, it is interesting
to point out that, while the first term in (\ref{Eq8}) represents
the usual relation between the angular momentum and the angular
velocity of a slowly rotating Kerr black hole, the second term on
the r.h.s of (\ref{Eq8}) is a direct consequence of the dragging of
inertial frames caused by the orbiting ring \cite{Will}.

A simple inspection of the compact expression (\ref{Eq8}) reveals
the physically important fact that, unlike vacuum Schwarzschild
black holes, a {\it zero} angular momentum ($J_{\text{H}}=0$) black
hole in the non-linearly coupled black-hole-orbiting-ring system is
characterized by the {\it non}-zero horizon angular velocity
\begin{equation}\label{Eq9}
\omega_{\text{H}}(J_{\text{H}}=0)={{2x^3}\over{M^3_{\text{ir}}}}\cdot
mj\ .
\end{equation}

In addition, it has been explicitly proved in \cite{Will} that, to
second order in the angular velocity of the black-hole horizon, the
composed axisymmetric black-hole-orbiting-ring system is
characterized by the total gravitational energy
\begin{eqnarray}\label{Eq10}
E_{\text{total}}(x)=m-m\Phi(x)+M_{\text{ir}}+
2M^3_{\text{ir}}\omega^2_{\text{H}}-\omega_{\text{H}}
mj\Psi(x)-{{m^2x}\over{2\pi
M_{\text{ir}}}}\ln\Big({{8M_{\text{ir}}\over{xr}}}\Big)\
\end{eqnarray}
as measured by asymptotic observers. Here we have used the
dimensionless radially dependent functions
\begin{equation}\label{Eq11}
\Phi(x)\equiv 1-{{1-2x}\over{(1-3x)^{1/2}}}\ \ \ ; \ \ \
\Psi(x)\equiv 12{{x^3-2x^4}\over{1-3x}}\  .
\end{equation}

The various terms in the energy expression (\ref{Eq10}), which
characterizes the composed black-hole-orbiting-ring system, have the
following physical interpretations \cite{Will}:
\begin{itemize}
\item{The first term in the energy expression (\ref{Eq10}) represents the proper mass of the ring.}
\item{In order to understand the physical meaning of the second term in the
energy expression (\ref{Eq10}), it is worth pointing out that, in
the small-$x$ regime (large ring radius, $R\gg M_{\text{ir}}$), this
term can be approximated by the compact expression [see Eqs.
(\ref{Eq6}), (\ref{Eq10}), and (\ref{Eq11})]
$-M_{\text{ir}}m/2R\cdot[1+O(M_{\text{ir}}/R)]$, which is simply the
sum of the potential and rotational Newtonian energies of the ring
in the background of the central compact object. Thus, this term
represents the leading order (linear in the mass $m$ of the ring)
interaction between the central black hole and the orbiting ring.}
\item{In order to understand the physical meaning of the third and fourth terms in the
energy expression (\ref{Eq10}), it is worth pointing out that a
slowly spinning bare (isolated) Kerr black hole is characterized by
the simple mass-angular-velocity relation
$M_{\text{Kerr}}=M_{\text{ir}}+2M^3_{\text{ir}}\omega^2_{\text{H}}+O(M^5_{\text{ir}}\omega^4_{\text{H}})$.
Thus, the third and fourth terms in (\ref{Eq10}) can be identified
as the contribution of the slowly spinning central black hole to the
total energy of the system. Interestingly, taking cognizance of Eq.
(\ref{Eq9}) one learns that due to the general relativistic frame
dragging effect, which is caused by the orbital motion of the ring,
the fourth term in (\ref{Eq10}) contains a self-interaction
contribution [of order $O(m^2/M_{\text{ir}})]$ to the total energy
of the composed black-hole-orbiting-ring system.}
\item{The fifth term in the energy expression (\ref{Eq10}) is a
non-linear spin-orbit interaction between the slowly spinning
central black hole and the orbiting ring. This energy term plays a
key role in our composed black-hole-orbiting-ring toy model system
since it is expected to mimic, at least qualitatively, the
physically analogous non-linear spin-orbit interaction in the
original black-hole-orbiting-particle system. Taking cognizance of
Eq. (\ref{Eq9}) one learns that, due to the intriguing general
relativistic phenomenon of frame dragging, the spin-orbit
interaction term in (\ref{Eq10}) contains a non-linear contribution
to the total energy of the composed black-hole-orbiting-ring system
which is of order $O(m^2/M_{\text{ir}})$.}
\item{The sixth term in the energy expression (\ref{Eq10}) is the
gravitational self-energy of the ring \cite{Tho} (not discussed in
\cite{Will}), where $r\ll R$ is the half-thickness of the ring. This
energy contribution represents the inner interactions between the
{\it many} particles that compose the axisymmetric ring. Since our
main goal in the present paper is to present a simple analytical
toy-model for the physically more interesting (and mathematically
more complex) two-body system in general relativity, which is
characterized by a {\it single} orbiting particle, we shall not
consider here this many-particle energy contribution. This
approximation allows one to focus the physical attention on the
general relativistic {\it frame-dragging} effect which characterizes
both the black-hole-orbiting-particle system and the
black-hole-orbiting-ring system.}
\end{itemize}

Taking cognizance of Eqs. (\ref{Eq7}), (\ref{Eq9}), (\ref{Eq10}),
and (\ref{Eq11}), one finds the compact functional expression
\begin{eqnarray}\label{Eq12}
E_{\text{total}}(x)=M_{\text{ir}}+m\cdot\Big[{{1-2x}\over{(1-3x)^{1/2}}}+
{{8x^5(-2+3x)}\over{(1-3x)^2}}\cdot\eta+O(\eta^2)\Big]\
\end{eqnarray}
for the total gravitational energy of the non-linearly coupled
black-hole-orbiting-ring system.

In the decoupling $R/M_{\text{ir}}\to\infty$ limit, in which the
ring is located at spatial infinity, the system is characterized by
the presence of two non-interacting physical objects: (1) a bare
(unperturbed) Schwarzschild black hole of mass $M=M_{\text{ir}}$
\cite{Notesmir}, and (2) a ring of proper mass $m$. Thus, the total
energy of the black-hole-ring system in the
$R/M_{\text{ir}}\to\infty$ limit is given by the simple expression
[see Eq. (\ref{Eq12}) with $x\to 0$]
\begin{eqnarray}\label{Eq13}
E_{\text{total}}(R/M_{\text{ir}}\to\infty)=M+m\ \ \ \text{with}\ \ \
M=M_{\text{ir}}\  .
\end{eqnarray}
Energy conservation implies that the marginally bound orbit of the
composed black-hole-orbiting-ring system is characterized by the
same total gravitational energy \cite{Notemmir}
\begin{eqnarray}\label{Eq14}
E_{\text{total}}(x=x_{\text{mb}})=M_{\text{ir}}+m\
\end{eqnarray}
as measured by asymptotic observers. Substituting the relation
(\ref{Eq14}) into Eq. (\ref{Eq12}), one finds the simple expression
\begin{equation}\label{Eq15}
x_{\text{mb}}={1\over
4}\cdot\Big[1+{{5}\over{16}}\cdot\eta+O(\eta^2)\Big]\
\end{equation}
for the $O(m/M_{\text{ir}})$-corrected location of the marginally
bound circular orbit in the composed black-hole-orbiting-ring
system.

Substituting the dimensionless radial coordinate (\ref{Eq15}) of the
marginally bound orbit into the functional expression \cite{Will}
\begin{equation}\label{Eq16}
M_{\text{ir}}\Omega=x^{3/2}\cdot\Big[1-4x^{3/2}\cdot M_{\text{ir}}\omega_{\text{H}}+O[(M_{\text{ir}}\omega_{\text{H}})^2]\Big]\
\end{equation}
for the dimensionless orbital frequency of the axisymmetric orbiting
ring and using Eqs. (\ref{Eq7}) and (\ref{Eq9}) \cite{Notesn}, one
obtains the $O(m/M_{\text{ir}})$-corrected expression
\begin{equation}\label{Eq17}
M_{\text{ir}}\Omega_{\text{mb}}={1\over
8}\cdot\Big[1+{{13}\over{32}}\cdot\eta+O(\eta^2)\Big]\
\end{equation}
for the characteristic orbital frequency of the marginally bound
circular geodesic in the composed black-hole-orbiting-ring system.


\section{Summary}

We have analyzed the physical and mathematical properties of a
composed black-hole-orbiting-ring system. In particular, we have
proposed to use this analytically solvable conservative
\cite{Notengw} system as a simple toy model for the conserved
dynamics of the astrophysically more interesting (and mathematically
more complex) black-hole-orbiting-particle system in general
relativity.

Our main goal was to provide a simple qualitative analytical
explanation for the {\it increase} in the orbital frequency of the
marginally bound circular geodesic that has recently been observed
numerically in the physically important work \cite{Barackmb}. To
this end, we have used the non-trivial spin-orbit interaction
between the central black hole and the orbiting ring, which is known
in a closed analytical form to second order in the dimensionless
angular velocity $M_{\text{ir}}\omega_{\text{H}}$ of the black-hole
horizon, in order to capture the essential physical features of a
similar non-linear spin-orbit interaction which is expected to
characterize the conservative dynamics of the
black-hole-orbiting-particle system.

Interestingly, the {\it analytically} derived expression [see Eqs.
(\ref{Eq2}) and (\ref{Eq17})]
\begin{equation}\label{Eq18}
{{\Delta\Omega_{\text{mb}}}\over{\Omega_{\text{mb}}}}={{13}\over{32}}\cdot
\eta+O(\eta^2)\
\end{equation}
for the dimensionless $O(m/M_{\text{ir}})$-shift in the orbital
frequency of the marginally bound circular geodesic in the composed
black-hole-orbiting-ring system provides the correct order of magnitude (with
the correct sign) for the corresponding shift in the orbital
frequency of the marginally bound circular geodesic of the
physically more interesting black-hole-orbiting-particle system.

This qualitative agreement suggests that the observed shift
(\ref{Eq3}) in the characteristic orbital frequency of the
marginally bound circular geodesic is mainly determined by the
general relativistic effect of dragging of inertial frames by
orbiting objects (the non-linear spin-orbit interaction between the
orbiting object and the generators of the central black-hole
horizon).

\bigskip
\noindent
{\bf ACKNOWLEDGMENTS}
\bigskip

This research is supported by the Carmel Science Foundation. I thank
Yael Oren, Arbel M. Ongo, Ayelet B. Lata, and Alona B. Tea for
stimulating discussions.


\begin{thebibliography}{99}
\bibitem{Noteirr} The irreducible mass of a black hole is related to its surface
area $A$ by the simple relation $M_{\text{ir}}=(A/16\pi)^{1/2}$. For
a spherically symmetric vacuum Schwarzschild black hole, the
irreducible mass coincides with the total ADM mass $M$ of the
spacetime: $M_{\text{ir}}=M$.
\bibitem{Notengw} It is worth stressing the fact that the dynamics
of the axisymmetric black-hole-orbiting-ring system is conservative in the
sense that, due to its simple symmetry, it contains no gravitational waves. Likewise, the
conservative part of the dynamics of the composed black-hole-orbiting-particle system
ignores the emission of energy and angular momentum in the form of gravitational waves.
\bibitem{Notesmir} Note that a bare Schwarzschild black hole is characterized by the simple relation
$M_{\text{ir}}=(A/16\pi)^{1/2}=M$.
\bibitem{Notemmir} We consider composed black-hole-orbiting-ring configurations
which are characterized by a fixed value of the geometrically
important quantity $M_{\text{ir}}$ (that is, a fixed value of the
central black-hole surface area).
\bibitem{Notesn} Note that one finds from Eqs. (\ref{Eq7}) and (\ref{Eq9}) the
dimensionless relation
$M_{\text{ir}}\omega_{\text{H}}={1\over8}\eta\cdot [1+O(\eta)]$ for
the angular velocity of the black-hole horizon when the orbiting
ring is in its marginally bound orbit (\ref{Eq15}).
\end{thebibliography}
\end{document}